\documentstyle[preprint,aps,floats]{revtex}
\begin{document}
\preprint{PUP-TH-1494 (1994)}
\title{  Electroweak Baryogenesis from a Classical Force}
\author{ Michael Joyce, Tomislav Prokopec, and Neil Turok}
\address{Joseph Henry Laboratories, Princeton University,
Princeton, NJ 08544}
\date{8/19/94}
\maketitle

\begin{abstract}
We describe a new effect which produces baryons
at a first order electroweak phase transition. It operates
when there is a CP-violating field present on propagating
bubble walls. The novel aspect
is that it involves a purely classical force, which alters
the motion of particles across the wall and through diffusion
creates a chiral asymmetry in front of the wall. We develop a
technique for computing the baryon asymmetry using
the Boltzmann equation, and a fluid approximation
which allows us to  model
strong scattering effects. The final formula for the
baryon asymmetry has a
remarkably simple form.
\\
\end{abstract}

The last ten years have seen a growing realisation that the
standard electroweak theory
satisfies Sakharov's conditions for baryogenesis:
$B$ violation, departure from thermal equilibrium,
$C$ and $CP$ violation \cite{Sakharov}.
Nonperturbative $B$ violating processes involving the
electroweak $SU(2)_L$ chiral anomaly, appear
unsupressed at high temperatures \cite{Ambjorn}.
The electroweak transition
is  weakly first order
for light Higgs masses, $M_H < 80$ GeV \cite{Farakos}.
It proceeds via
bubble nucleation, with departures from thermal
equilibrium on and around  bubble walls \cite{DineMcl}.
Finally,
$C$ and $CP$ are violated, the latter via the phase
in the KM matrix.
The $CP$ violation in the minimal theory is very small,
but there may be amplification mechanisms which
enhance it, or additional Higgs fields
with $CP$ violation in the Higgs potential
(for reviews see  \cite{Turokreview}, \cite{CKNreview}).

In this Letter we
study baryogenesis due to a $CP$ violating condensate on
bubble walls.
Several mechanisms have already been pointed out
 through which
such a condensate can produce a baryon asymmetry.
It couples via a term in the effective action to
bias the winding of the gauge and Higgs fields
\cite{TurokZadrozny}.
It also acts to bias hypercharge violating particle
interactions, which in a certain constrained thermal
equilibrium
favors baryon production - `spontaneous'
baryogenesis
\cite{CKNspontaneous}, \cite{DT},
\cite{JPThypercharge}, \cite{JPTthickwalls}.

In addition to these {\it local} effects, particle transport
can carry the $CP$ violation present on the
wall into the unbroken phase, where the $B$ violation
rate is maximal \cite{CKNtops}.
Until recently, this {\it nonlocal}
barogenesis  was thought to necessarily involve
quantum mechanical
interference - the idea was that the condensate causes
$CP$ violation in particle-wall scattering amplitudes,
leading to a chiral flux (i.e. more $t_L$'s than
$\overline{t_L}$'s) being injected into the unbroken
phase where it drives baryon production.
Top quarks are the obvious candidate
because of their large mass and thus strong coupling to
the wall.
However significant asymmetries can only be produced
for very thin walls
because the quantum interference effect tends to be
destroyed by
the strong (QCD) scattering of the quarks, and is also
WKB suppressed for walls much thicker than the inverse
top mass. Partly for these reasons
we  considered tau leptons as an alternative
because they are much more weakly
coupled to the plasma
\cite{JPTleptons}, \cite{JPTthinwalls}.

In this Letter we discuss a new, purely
classical mechanism through which nonlocal
baryogenesis can be driven. It does not
rely on quantum mechanical interference, and thus may be
calculated from a Boltzmann equation, which
we shall solve analytically in a fluid approximation.
The physical picture is extremely simple:
the classical force
drags an excess of chiral charge onto the wall, leaving
a compensating deficit of chiral charge in front of the wall,
which drives baryogenesis.
Particle transport is the key to this
mechanism --- if particles
are free to diffuse in the medium, they are
free to respond to
the chiral force on the wall. Conversely, if
particles cannot
move relative to the plasma (e.g. top quarks in the
limit of large $\alpha_S$) the whole effect goes away, and
only local baryogenesis is possible.

The qualitative criterion for efficient transport is that
a particle should be able to diffuse a distance $x>L$,
the wall width,
in the time the wall takes to pass $t =L/v_w$, with $v_w$
the wall velocity. Setting $x^2 \sim Dt$, with $D$
the diffusion
constant, we find

$\bullet$ {\bf Condition 1.}  $v_w < D/L$
(for efficient transport)

Secondly we require that the phase space
density approach a local thermal equilibrium (LTE) form, in
which a chiral charge builds up on the wall. This requires
that the equilibration time $\tau$ be smaller than the time
of passage of the wall $L/v_w$.
We will discuss below how we determine this timescale
$\tau$ within our calculation and  find
that $\tau \sim 3D$. We therefore require

$\bullet$ {\bf Condition 2.}  $v_w <  L/ 3D$
(for LTE to establish)

And finally we demand that the semiclassical (WKB)
approximation must be accurate. For this to be true,
the effect should come from particles with typical momenta
$p_z$ such that

$\bullet$ {\bf Condition 3.}  $ |p_z|  >> L^{-1}$
(for classicality)

Conditions 1-3 are plausibly met in the standard
electroweak model,
and minimal extensions.
As long as they are, we shall see that the
final baryon asymmetry is to a good approximation
independent of
$v_w$, $L$, and $D$.

One loop calculations
yield  $L\sim 20-40 T^{-1}$\cite{DineMcl},
and in our mechanism the particles
that dominate the effect have $|p_z| \sim T$,
so Condition 3 is easily
satisfied. We have estimated \cite{JPTthickwalls}
$D \sim 5 T^{-1}$ for quarks.
Calculations of the wall velocity are difficult
\cite{DineMcl},
but indicate velocities in the range $v_w \sim .05-1$.
Conditions 1 and 2 are therefore  satisfied
for a large part of the parameter space indicated by
studies of the phase transition.
However, we expect particle diffusion in front of the wall
to become negligible
if $v_w > v_s$, the speed
of sound in the plasma, in which case {\it local}
baryogenesis should
dominate.

Recently, we pointed out in \cite{JPThypercharge}
that transport effects could be important in
`spontaneous'
baryogenesis \cite{CKNspontaneous}, and raised doubts about
this mechanism because transport effects spoil
the constraints
imposed in local equilibrium
calculations \cite{JPThypercharge},
unless the walls are very thick.
Subsequently, one of us
pointed out that transport
phenomena could actually enhance this
mechanism \cite{JoyceSintra}
and this has been independently
explored using a diffusion equation
in \cite{CKNnew}. We believe that our procedure
provides a more complete framework within
which both this and
the classical force effect we focus on here,
can be computed and give a detailed treatment of
both effects
in \cite{JPTthickwalls}.

We begin with a derivation of the chiral force.
The Lagrangian describing a fermion
moving in the classical background of a bubble wall
with $CP$ violating condensate is
\begin{equation}
{\cal L}_{\it ch}=\overline{\Psi}\gamma^\mu i
(\partial_\mu
- i g_A
\tilde{Z}_\mu\gamma^5)\Psi
- m \overline{\Psi}\Psi
\label{eq:chirallagrangian}
\end{equation}
 where $g_A \tilde{Z}_\mu = g_A Z_\mu^{GI}
-{1\over 2} {v_2^2 \over v_1^2 +v_2^2}
\partial_\mu \theta$,
$g_A= {1\over 4} \sqrt{g_1^2+g_2^2}$ for $t$ quarks
\cite{JPTthinwalls}.
The two contributions
the $CP$-odd scalar
field $\theta$ which is the relative phase of the
two Higgs fields in a
two Higgs theory,
$\varphi^\dagger_2 \varphi_1 = R e^{i\theta}$, and
the $Z_{\mu}$ condensate discussed in \cite{NasserTurok},
which may be present even in the minimal theory.
The notation $GI$ implies that this is the gauge invariant
combination of the gauge fields and Higgs phases which
diagonalises the Higgs kinetic terms.
We treat the wall as planar, and assume it has
reached a stationary
state in which the Higgs and gauge fields are
functions of $z-v_wt$.
In this case, the field $\tilde{Z}_\mu$ is pure gauge.

The axial coupling in
(\ref{eq:chirallagrangian}) leads to a classical
chiral force
as follows. In the rest frame of the wall
$\tilde{Z}_\mu = (0,0,0, \tilde{Z}_z(z))$.
{}From the corresponding Dirac equation,
setting $\psi \propto e^{-i p.x} $, one finds the
following dispersion
relation \cite{JPTthickwalls}
\begin{equation}
E = \left [ p_\perp ^2 +\left (\sqrt{p_z^2+m^2}\mp
g_A \tilde{Z}_z \right )^2 \right ]
 ^{\frac {1}{2}} \quad S_z = \pm{1 \over 2}
\label{eq:dispersionrelation}
\end{equation}
for both particles and antiparticles.
$S_z$ is the component of the spin in the $z$ direction,
measured in the frame in which $p_\perp$ vanishes.
In the WKB approximation, this dispersion relation
accurately describes particles as they move  across
a bubble wall - the local eigenstates in
(\ref{eq:dispersionrelation}) shall form the basis of our
treatment.
The particles we are most interested in for
baryogenesis are
left handed particles (e.g. $t_L$'s),
and their antiparticles ($\overline{t_L}$'s, which are
right handed), because these couple to the chiral anomaly.
For large $|p_z|$, these are easily identifiable in terms
of the eigenstates in
(\ref{eq:dispersionrelation}).
Note that they couple {\it oppositely}
to the $\tilde{Z}$ field.

The group velocity of a WKB wavepacket is
determined from the dispersion
relation by
$v_z = \dot z = \partial E / \partial p_z$, and
energy conservation $\dot E =0= \dot z (\partial E /
\partial z) + \dot p_z (\partial E / \partial p_z)$
then implies that
$\dot p_z = -\partial_z E$.  These are of course Hamilton's
equations for the motion of a particle. From these it
is straightforward to calculate the acceleration
\begin{equation}
{d v_z \over dt} = -{1\over 2}
{(m^2)'\over E^2} \pm{(g_A \tilde{Z}_z m^2)'
\over E^2 \sqrt{E^2-p_\perp^2} }  +o(\tilde{Z}_z^2)
\label{eq:acceleration}
\end{equation}
where $E$ and $p_\perp$ are constants of motion.

This chiral force provided by the $\tilde{Z}_z$ field
effectively produces a potential well which draws
an excess chiral charge onto the wall, and leads
to a compensating
deficit in a  `diffusion tail' in front of the wall.
There is net baryon production because $B$ violation
is suppressed on the bubble wall.

We now seek to describe the particle excitations with
dispersion relations
(\ref{eq:dispersionrelation}) as
classical fluids.
We focus on
particles with large  $|p_z|\sim T>>m$ for three reasons:
they dominate phase space,  the
$WKB$ approximation is valid, and
the dispersion relation simplifies so one can
identify approximate chiral eigenstates.
The $S_z=+{1\over2}$, $p_z<0$
branch, and the $S_z=-{1\over 2}$, $p_z>0$ branch
constitute one, approximately left
handed fluid $L$, and the other two branches
an approximately  right handed
fluid $R$.

The Boltzmann equation is:
\begin{equation}
d_t f \equiv \partial_t f + \dot z \partial_z f + \dot p_z
\partial_{p_z} f = - C(f)
\label{eq:boltzmannequation}
\end{equation}
where $\dot z$ and $\dot p_z$ are calculated as above,
and $C(f)$
is the collision integral.
This can in principle be solved fully. However
to make it analytically tractable we truncate
it with a fluid
approximation which we now discuss.
When a collision rate is large, the collision integral
forces the distribution functions towards the
local equilibrium form
\begin{equation}
f={1\over {{\rm e}^{\beta [\gamma (E-v p_z)-\mu]}+1}}
\label{eq:distributionfunction}
\end{equation}
where $T= \beta^{-1}$,
$v$ and $\mu$ are functions of $z$ and $t$, and
$\gamma=1/(1-v^2)^{1/2}$.
These parametrise the fluid velocity  $v$, number
density  $n$ and energy density $\rho$.
We are going to treat the approximately left-handed
excitations $L$  and their antiparticles $\overline{L}$
as two fluids, making an {\it Ansatz}
 of the form (\ref{eq:distributionfunction}) for each.

As mentioned, the {\it Ansatz}
(\ref{eq:distributionfunction}) does allow
us to describe perturbations in the energy density,
number density
and velocity of each fluid, and we expect it
to give a reasonable
qualitative description of the true phase
space density perturbations.
As far as the temperature and velocity
perturbations are concerned, we
probably cannot expect this form to be more
than qualitatively correct,
because
the dominant interactions which
establish thermal equilibrium
are
those
with the background plasma, which are also
responsible for setting $\delta v$
and $\delta T$ to zero.
This is not
true however of the chemical potential perturbation,
which is
only attenuated by slower chirality changing processes.
So as long as we can check that
$\delta T/T$ and $\delta v$ are small,
compared to $\mu/T$, we believe that
(\ref{eq:distributionfunction}) should actually provide
an accurate parametrisation of the phase space density.

The collision integrals are
evaluated in the
approximation that the particle interactions are local,
by using the Dirac spinors appropriate to
the local value of $m$ and $\tilde{Z}_z$,
taken to be constant.
This is reasonable for the two body scattering effects we
consider, because the QCD interactions are short
ranged, the  Debye screening
length $m_{gluon}^{-1}$ being smaller than $ T^{-1} $ and
very much smaller than $L$.

The ansatz (\ref{eq:distributionfunction})
has three arbitrary
functions, which are fully determined from
three independent
moments of the Boltzmann equation:
we take ${\int d\!\!\!^-}^3p$,
$\int{d\!\!\!^-}^3 p E$, and $\int {d\!\!\!^-}^3 p p_z$.
For
a single interacting fluid these yield the
continuity, energy and momentum equations.
We are interested in the chiral density, the
difference between $L$ and $\overline{L}$  chemical
potentials, since this quantity drives the baryon asymmetry.
We work to first order in $\tilde{Z}_z$
and $v_w$.

The fluid equations  for particle minus
antiparticle perturbations ($\delta T \equiv \delta T(L)
-\delta T(\,\overline{L}\,)$,
$\mu =\mu(L)-\mu(\,\overline{L}\,)$,
 $\delta v =\delta v(L) -\delta v(\,\overline{L}\,)$) are,
 in the rest frame of the wall
\begin{eqnarray}
-v_w {\delta T^\prime\over T_0}
+\frac {1} {3} \delta v^\prime
-a v_w {\hat\mu^\prime\over T_0} &&=
-\overline{\Gamma}_\mu ( \frac{\hat \mu}{T_0})
-{\Gamma}_\mu (\frac{\Delta}{T_o} )
\label{eq:fluidequationi} \\
-v_w{\delta T^\prime \over T_0} +\frac {1} {3}
\delta v^\prime -v_w b {\hat\mu^\prime\over T_0
}
 &&=
-\Gamma_T  {\delta T\over T_0}
-\overline{\Gamma}^*_\mu (\frac{\hat \mu}{T_0})
-\Gamma^*_\mu ..
\label{eq:fluidequationii}  \\
{\delta T^\prime\over T_0} +
 b {\hat\mu^\prime\over T_0}
- 2 c v_w &&{(g_A \tilde{Z}_z m^2)^\prime \over T_0^3}
=
-\Gamma_v \delta v
\label{eq:fluidequationiii}
\end{eqnarray}
where the shifted chemical potential difference is
$\hat\mu=\mu-2v g_A \tilde{Z}_z$,
$(\hat \mu)$ denotes the signed sum of
chemical potentials for particles
participating in the reaction,
$\Delta = (\mu) =(\hat \mu -2 v_w g_A \tilde{Z}_z)$,
is the difference
between shifted $L$ and $\overline{L}$ potentials,
and
prime denotes $\partial_z$,
$a=\pi^2/27\zeta_3$, $b=n_0 T_0/ \rho_0$,
 $c=\ln 2/14 \zeta_4$, $\zeta_4=\pi^4/90$,
$n_0=3\zeta_3 T_0^3/4\pi^2$,
$\rho_0=21\zeta_4 T_0^4/8\pi^2$
 and $\zeta$  is the Riemann
$\zeta$-function. The derivation of
these equations is simplest if one  shifts
the canonical momentum to $k_z=p_z+g_A \tilde{Z}_z$
and the chemical potential to $\hat \mu$.
In this way the correct
massless limit emerges as one expands
in powers of $\tilde{Z}_z$.
The relevant collision integrals may be calculated
at zero background fields \cite{footnote2}.
$\Gamma_v$ is simply related to the diffusion
constant $D$ -
it is easily seen that
$D=(n_0 T_0/4a \rho_0) \Gamma_v^{-1}\approx \frac{1}{4}
\Gamma_v^{-1}$. In fact we  find $\Gamma_T \approx
{1\over 3} \Gamma_v, \Gamma_v \approx T/20$
\cite{JPTthickwalls}.

$\overline{\Gamma}_\mu$ and $\overline{\Gamma}_\mu^*$
are
derived from hypercharge {\it conserving}
chirality flip processes, such as those
involving external Higgs particles.
In this case, the $\tilde{Z}_z$
contribution to sum of chemical potentials
vanishes. $\Gamma_\mu$ and $\Gamma_\mu^*$
are the rates for hypercharge {\it violating}
chirality flip processes, which are  $m^2$
suppressed and for these the
$\tilde{Z}_z$ contribution does not cancel.
These latter are the
terms driving `spontaneous' baryogenesis, in its
new `diffusion-enhanced' form.
This enhancement was mentioned
in a talk by one of us \cite{JoyceSintra}, and
explored independently and in much
greater detail by Cohen et. al.
\cite{CKNnew}. In the formalism represented
by the above equations,
the `spontaneous baryogenesis' and `classical force'
driving
terms are both included - this is fully discussed
in ref.
\cite{JPTthickwalls}. Here we focus on
classical  force term,
which since the equations are linear
can be considered independently
of the `spontaneous baryogenesis' source terms.

We proceed to solve equations
(\ref{eq:fluidequationiii}) to find the perturbations
produced by the force term. We simplify by setting
$\Gamma_\mu$,
$\Gamma^*_\mu$,
$\overline{\Gamma}_\mu$, and
$\overline{\Gamma}^*_\mu$ equal to zero.
We show in \cite{JPTthickwalls} that the suppression
they produce of the classical force  effect is,
for a large range of parameters, a
factor between ${1\over 2}$
and 1.
With this simplification we can derive our result
in a few lines.
First, from
(\ref{eq:fluidequationii}) we see that if $v_w D/ L<1$
the temperature fluctuation is smaller
than the velocity
or chemical potentials by this  factor
(using $ \Gamma_T \approx 1/12D$).
This explains how we arrived at our
Condition 2 above.
Then from
(\ref{eq:fluidequationi})
we find a simple relation $\delta v \sim
v_w \mu$. So we do indeed find the
temperature and velocity
perturbations are small.
The {\it reason} the velocity perturbation
is small is quite general - from
the continuity equation it follows that the
velocity perturbation
required
to create a given chiral excess in the wall frame
is proportional to the wall velocity.
This should  to be true independently of the
detailed form of the phase space density,
which as mentioned
before we cannot expect to be exact.
Since $\delta v$ is small,
we can  drop the r.h.s. of
(\ref{eq:fluidequationiii})
because on the wall it is of order
$ v_w L / D$ compared to
the $\hat \mu$ term, which is small by Condition 1.

We are left with a relation between the chemical
potential $\hat \mu$ and the force term on the wall,
from
(\ref{eq:fluidequationiii}):
\begin{equation}
 {\hat\mu} = -\frac{2 \ln 2}{3\zeta_3}
v_w {g_A \tilde{Z}_z m^2
 \over T_0^2}
\qquad {\rm on} \,\, {\rm the} \,\, {\rm wall}
\label{eq:shiftedmu}
\end{equation}
We can now determine
$\hat\mu$ in front of the wall as follows.
Integrating (\ref{eq:fluidequationii}) and
(\ref{eq:fluidequationiii})
(with all $\Gamma_\mu$'s zero),
gives
$\int_{-\infty}^\infty \delta T \approx 0$
and $\int_{-\infty}^\infty
 \delta v=0$.
Then
integrating
(\ref{eq:fluidequationi}) twice we
find  $\int_{-\infty}^\infty \mu= 0$
{\it i.e.} no net integrated chemical potential
perturbation is
generated. This means that the chemical potential
generated on
the wall is compensated by an opposite
chemical potential
off it. As mentioned above, off the
wall the equations for $\mu$ reduce to
the diffusion equation,
and it is straightforward to see that in the absence of
particle number violation the only nontrivial
solution for $\mu$
is a diffusion tail {\it in front} of the wall.
This is where
the chiral charge deficit occurs, which
drives baryogenesis.

Hence the
integral of
the chemical potential in front of the
wall $(z>0)$ equals:
\begin{equation}
\int_{0}^{\infty}d z \mu =
\frac{2 \ln 2}{3\zeta_3} v_w
\int_{\rm wall} dz  g_A \tilde{Z}_z m^2
\label{eq:integralmu}
\end{equation}

Now, using
the standard formula for baryon number violation
\begin{equation}
\dot n_B=-v_w n_B^\prime = - {3\over 2}
N_C {\bar\Gamma_s\over T^3}
(\mu_{t_L} - \mu_{\overline{t}_L})
\label{eq:baryonrate}
\end{equation}
where $\bar\Gamma_s=\kappa (\alpha_w T)^4$
is the weak sphaleron
rate in the unbroken phase, $N_c$ is the number of colors,
$\kappa\in [0.1, 1]$ \cite{Ambjorn}.
We have re-expressed $\mu$ in terms
of top quark and antiquark chemical potentials.
We arrive at a  formula for the baryon to entropy ratio:
\begin{equation}
\frac {n_B}{s}= {135 \ln 2 \over 2 \pi^2\zeta_3}
{\kappa \alpha_w^4 \over
g_*} \int dz {m^2 g_A \tilde{Z}_z \over T^2}
\approx 4
{\kappa \alpha_w^4 \over
g_*} \int dz {m_t^2 g_A \tilde{Z}_z \over T^2}
\label{eq:baryontoentropy}
\end{equation}
where $s=\frac{2\pi^2}{45} g_* T^3$ is the
entropy density,
with $g_*\approx 100$  the effective number of
degrees of freedom.

This result is remarkably simple
- all dependence on the wall velocity,
thickness and the diffusion constant drops
out, provided
Conditions 1 and 2
are satisfied.
It is also quite large:  $(m_t/T)^2 \sim 1$, so
$n_B/s \sim 4 \times 10^{-8} \kappa \theta_{CP}$ where
$\theta_{CP}$ characterises the strength
of the $CP$ violation.
In a longer paper \cite{JPTthickwalls} we give
a more detailed derivation
of (\ref{eq:baryontoentropy}) with a full  discussion
of parameter dependences, including the effect
of the $\Gamma_\mu$
terms we have neglected here.

The $m^2$ dependence in (\ref{eq:baryontoentropy})
means that,
at least with standard model-like Yukawa couplings,
the top quark dominates the effect.
The
mass-over-temperature suppression can
be significant, if the $\tilde{Z}_z$ field is
localised
on the front of the wall where the Higgs vev is small.

The calculation of the classical force effect above uses
the opposite (WKB) approximation to those employed in
quantum mechanical reflection calculations (thin walls)
\cite{CKNtops}, \cite{JPTleptons}).
The classical
force calculation is in some respects `cleaner', because
the production of chiral charge and its
diffusion are treated
together.
The classical force affects particles from
all parts of the spectrum, mostly with typical energies $E\sim
T$, and  with no preferential direction, while the quantum
mechanical
effect comes mainly from
 particles with a very definite ingoing momentum
perpendicular to the wall: $p_z\approx m_H$ (Higgs mass).
The quantum result falls off strongly with $L$
(at least as $L^{-2}$) as the $WKB$ approximation becomes good.
The quantum result also has a $v_w^{-1}$ dependence
coming from the diffusion time in the medium, which
the classical result loses because the force term is
proportional to  $v_w$.

Finally, we mention possible extensions of these
methods. In the above treatment
we have comletely ignored collective plasma effects
- Debye screening,
Landau damping etc, and merely treated
{\it local} particle
interactions. The Boltzmann equation is
easily modified to
include these effects, with force terms due to
the electric (and magnetic) fields, which are solved for
self-consistently. The fluid truncation may be
a useful way
to compute the bubble wall velocity
(at least the friction
due to top quarks), and we shall return to
this in future work.
We intend also to extend these methods
to study $Z$-condensation
in the standard model \cite{NasserTurok}
in the presence of strong interactions.

{\it Acknowledgements.\/}
M.J is supported by a Charlotte Elizabeth Procter
Fellowship,
and the work of T.P.  and N.T.
is partially   supported by
NSF contract PHY90-21984, and the David and Lucile
Packard
Foundation.



\begin{thebibliography}{99}
\bibitem{footnote2}{
The collision integrals vanish identically
for arbitrary background fields at zero
$\delta T$, and $\delta v$
respectively, so they take the form $ (\delta T)
\Gamma(Z_z=0,m^2=0)(1 + o(Z_z) + o(m^2) +...)$
But any rate on the R.H.S. multiplies
a perturbation, of
$o(m^2 Z_z)$ so that the linear correction to the rate
contributes at  second order in $Z_z$ and can
be ignored.}
\end{thebibliography}
\end{document}